\documentstyle[epsf]{mn}

\newcommand{\etal}{et~al.}

\newcommand{\ionhy}{H{\sc ii}}

\newcommand{\water}{$\mbox{H}_{2}\mbox{O}$}
\newcommand{\methanol}{$\mbox{CH}_{3}\mbox{OH}$}
\newcommand{\formaldehyde}{$\mbox{H}_{2}\mbox{CO}$}

\newcommand{\transa}{$5_{1}\rightarrow6_{0}\mbox{~A}^{+}$}

\newcommand{\kms}{$\mbox{km~s}^{-1}$}

\newcommand{\micron}{\mbox{$\umu$m}}
\newcommand{\lta}{\raisebox{-0.6ex}{$\,\stackrel
{\raisebox{-.2ex}{$\textstyle <$}}{\sim}\,$}}
   
\def\epsout #1 {
  \centering
     \leavevmode\epsffile{./#1}
}
\newcommand{\dfig}[2]           
{
  \begin{center}
    \hfill
    \begin{minipage}[t]{0.42\textwidth}
       \epsfxsize=0.99\textwidth
       \epsout #1.eps
    \end{minipage}
    \hfill
    \begin{minipage}[t]{0.42\textwidth}
       \epsfxsize=0.99\textwidth
       \epsout #2.eps
    \end{minipage} \\
  \end{center}
}

\ifCUPmtlplainloaded \else
  \def\umu{\mu}

\fi

\title[A Survey for 6.7-GHz Methanol Masers I]
      {A Survey of the Galactic Plane for 6.7-GHz Methanol Masers I:
      $l$ = 325\degr -- 335\degr\/ ; $b$ = -0\fdg53 -- 0\fdg53}
\author[S.P. Ellingsen \etal]
       {S.P. Ellingsen$^{1}$, 
	M.L. von Bibra$^{1}$, 
	P.M. McCulloch$^{1}$,\cr
	R.P. Norris$^{2}$,
        A.A. Deshpande$^{1}$ \thanks{On leave from Raman
                                     Research Institute,
                                     Bangalore 560080, India}
        and  C.J. Phillips$^{1}$ \\
	$^{1}$Physics Department, University of Tasmania, GPO Box 252C, 
	      Hobart, TAS 7001.\\
	$^{2}$Australia Telescope National Facility, CSIRO, PO Box 76,
	      Epping, NSW 2121.}
\date{Received dd Month Year; in original form dd Month Year}
\pagerange{\pageref{firstpage}--\pageref{lastpage}}
\pubyear{1995}

\begin{document}

\label{firstpage}

\maketitle

\begin{abstract}
We report the results of the first complete survey of an area of the
Galactic Plane for maser emission from the 6.7-GHz \transa\/
transition of \methanol\/.  The survey covers a 10.6-square-degree
region of the Galactic Plane in the longitude range 325\degr -- 335\degr\/
and latitude range -0\fdg53 -- 0\fdg53.  The survey is sensitive to
masers with a peak flux density greater than $\sim$ 2.6~Jy.  The
weakest maser detected has a peak flux density of 2.3~Jy and the
strongest a peak flux density of 425~Jy.  We detected a total of 50
distinct masers, 26 of which are new detections.  We show that many
6.7-GHz \methanol\/ masers are not associated with {\em IRAS} sources,
and that some are associated with sources that have colours differing
from those of a typical ultra-compact \ionhy\/ region (UC\ionhy).  We
estimate that the number of UC\ionhy\/ regions in the Galaxy is
significantly more than suggested by {\em IRAS}-based estimates,
possibly by more than a factor of two.
\end{abstract}


\section{Introduction}
In the four years since the discovery of maser emission from the
\transa\/ transition of \methanol\/ several searches for 6.7-GHz
\methanol\/ maser emission have been made towards sites of OH and
12.2-GHz \methanol\/ masers \cite{Me1991b,Ma1992b,Ma1992a,Ca1995a}.
Schutte \etal\/ \shortcite{Sc1993} have searched toward sources
believed to be ultra-compact \ionhy\/ regions on the basis of their
colours in the {\em IRAS} (Infrared Astronomy Satellite) Point-Source
Catalog \shortcite{IRAS}.  In nearly all cases, the 6.7-GHz \methanol\/
masers are stronger than their 12.2-GHz and OH counterparts.  Thus,
the first search method is likely to find few weak 6.7-GHz masers.
The {\em IRAS} satellite had a beamwidth of 2~arcmin at 100~\micron\/,
so that in the extremely crowded and confused regions close to the
Galactic Plane many ultra-compact \ionhy\/ (UC\ionhy) regions may not
have been detected.  Thus, the second search method is likely to miss
a significant number of \methanol\/ masers associated with UC\ionhy\/
regions.

Here we present an untargeted search of a region of the Galactic Plane
known to be rich in masers of other molecular transitions.  An
untargeted search enables us to detect new 6.7-GHz masers associated
with sites of massive star formation and also to find any masers which
may be associated with different classes of object.  For several
reasons, the 6.7-GHz transition of \methanol\/ is very good for
detecting new sites of massive star formation.  It is the second
strongest maser transition known, after the 22-GHz transition of
\water, and as it occurs at a far lower frequency, the telescope
beam will be larger, enabling a more rapid search of a large portion
of the Galactic Plane.  In addition, the 6.7-GHz \methanol\/ masers
are less variable than either \water\/ or 12.2-GHz \methanol\/ masers
\cite{Ca1995b}.  Thus there is less chance that a source will drop
below the detectability threshold in the interval between the initial
survey and final observations.

The various transitions of \methanol\/ masers have been separated into
two classes on the basis of several observed properties \cite{Ba1987}.
The original distinction between the two classes of \methanol\/ masers
was that class II masers are typically associated with UC\ionhy\/
regions, which often also show OH and \water\/ maser emission.  The
6.7- and 12.2-GHz transitions are the most common examples of class II
\methanol\/ masers.  Class I \methanol\/ masers are usually found
offset from UC\ionhy\/ regions, and are typically weaker and have
fewer components than the class II masers.  Class II transitions,
particularly the 12.2-GHz transition, are often seen in absorption
toward the sites of class I \methanol\/ maser emission.  Recent
observations by Slysh \etal\/ detected 44.1-GHz class I \methanol\/
masers toward many 6.7-GHz class II \methanol\/ masers.  They found
that the strong 44.1-GHz masers were usually associated with weaker
6.7-GHz masers and that while the two transitions often shared the
same velocity range, there was no detailed correspondence between
features in the 44.1-GHz spectrum and those in the 6.7-GHz spectrum.

Since the release of the {\em IRAS} Point-Source Catalog
\shortcite{IRAS}, many searches for masers have been made by observing
towards {\em IRAS} sources which satisfy various criteria
\cite{Br1987b,Co1988,Pa1991,Sc1993}.  The {\em IRAS} satellite made
observations in four wavelength bands at 12, 25, 60 and 100~\micron.
Each flux density measurement has an associated quality flag, indicating
whether the observation is of high or moderate quality, or only an
upper limit.  The most frequently used criteria for selecting
UC\ionhy\/ regions from the {\em IRAS} catalog were developed by Wood
\& Churchwell \shortcite{Wo1989}.  They found that {\em IRAS} sources
which satisfied $Log_{10}(S_{60}/S_{12}) \geq
1.30$~\&~$Log_{10}(S_{25}/S_{12}) \geq 0.57$ were more likely to be
UC\ionhy\/ regions than those in other regions of the colour--colour
diagram.  Currently, searches using {\em IRAS} selection criteria seem
to be amongst the most reliable methods of finding UC\ionhy\/ regions.
However, if all 6.7-GHz \methanol\/ masers are associated with
UC\ionhy\/ regions, they may allow a more accurate determination of
the number of UC\ionhy\/ regions in the Galaxy.

\section{Observations}

The observations were made between 1993~April and 1994~September using
the University of Tasmania's 26-m antenna at the Mt Pleasant
Observatory.  This antenna has rms pointing errors of 0.7~arcmin and a
7-arcmin HPBW at 6.7~GHz.  A dual-channel, cryogenically cooled HEMT
receiver, with two orthogonal circular polarizations, was used for all
observations.  A 1-bit digital autocorrelation spectrometer, with
512~channels per polarization covering 2.5~MHz, was used for the
survey observations.  This configuration gives a velocity coverage of
112.5~\kms\/ and a velocity resolution after Hanning smoothing of
0.44~\kms\/ at 6668.518~MHz, the rest frequency of the \transa\/
transition of \methanol\/.  Between Galactic longitudes 325\degr\/ and
330\degr\/, observations were centred on a velocity of -70~\kms\/ with
respect to the local standard of rest, and between 330\degr\/ and
335\degr\/ observations were centred on a velocity of -30~\kms\/.
Hydra A and Virgo A were used as flux density calibrators, with
assumed peak fluxes of 10.4 and 54.1~Jy in a 7-arcmin beam.  The
survey observations were made in an equilateral triangular grid
pattern, with each grid point separated by 3.5~arcmin (half the HPBW
at the observing frequency) from all adjacent points.  In total, the
survey consisted of observations at approximately 3500 sky positions.
The on-source integration time at each grid point was 10~min.  The
system noise was typically $<$ 650~Jy, yielding an rms noise level of
330~mJy in each spectral channel after Hanning smoothing and averaging
the two polarizations.

As the survey observations were made in a wide variety of conditions,
the sensitivity limit is not uniform across the entire region.  We
minimised variations in the sensitivity limit of the individual
spectra by limiting the range of elevations over which we observed,
and then we re-observed poor-quality spectra.  As maser profiles are
generally very narrow, high spectral-resolution observations are
required to measure the full peak flux density.  The attenuation of
the maser peak depends upon the width of the maser, the spectral
resolution, and the position of the maser with respect to the centre
of the spectral channel.  To determine a typical halfwidth for the
6.7-GHz \methanol\/ masers, we did a Gaussian component analysis of 17
of the sources in our sample.  Many of the features in the maser
spectra are clearly blended, significantly increasing the difficulty
of the Gaussian analysis.  To avoid errors introduced by our analysis
we used only the strongest three components from each of the spectra,
or the strongest only in the case of spectra with three or fewer
components.  For these 47 features, the mean width at the half-power
points of the 6.7-GHz \methanol\/ masers was 0.47~\kms.  The peak flux
density of a maser of width 0.47~\kms\/ would be reduced, on average,
to 68~per cent of its true value by observations made with the
spectral resolution of this survey.  As the maser will not usually be
at the centre of the telescope beam, the detected maser peak flux
density will be further attenuated typically by about 5.5~per cent.
With an rms level of 330~mJy, our effective 5-$\sigma$ detection limit
lies in the range 2.4 (best) to 3.0~Jy (worst), with a mean of 2.6~Jy.


Most of the masers were detected in several beam positions in the
initial survey.  The relative intensities were then used to obtain an
approximate position for the source.  We then determined a more
accurate position by observing a grid of 5~points centred on the
approximate position (1 at the approximate position and 4 in a square
surrounding it).  Some of the sources have positions determined from
the Parkes telescope, or Australia Telescope Compact Array (ATCA)
observations, and the rms difference between these and our positions
is $\sim$ 0.6~arcmin.  Finally, we made high signal-to-noise ratio
observations of all detected sources.  On-source integration times
ranged from 10 to 90 min, resulting in spectra with signal-to-noise
ratios of at least 15:1.  All but the weakest sources (326.40+0.51,
327.61-0.11 and 332.33-0.44) were re-observed using a correlator
configuration of two 512-channel spectra, each spanning 0.625 MHz,
resulting in a velocity resolution after Hanning smoothing of
0.11~\kms.  The three weakest sources were re-observed using the same
correlator configuration as for the main part of the survey.  Most
sources were observed at the positions which we determined, although
for several close groups of sources the Parkes and ATCA positions were
used because of the difficulty in determining exact separations with
our larger beam.  The position at which each source was observed is
listed in Table~\ref{tab:main}.

\section{Results}

These observations resulted in the detection of 50 individual 6.7-GHz
\methanol\/ masers (summarized in Table~\ref{tab:main}).  Spectra from
all sources are shown in Fig.~\ref{fig:spectra}.  Of the 50 sources,
37 lie in the spatial and velocity range which the survey sampled
completely.  The remaining 13 sources were detected either because
they had a sufficiently high flux density that they were detected even
though they lay outside the survey region, or because they were
detected by accident while refining source positions or taking
off-source reference spectra.  As these positioning observations were
centred at the detected velocity, they occasionally resulted in the
serendipitous discovery of sources offset in either position or
velocity from the parameter space sampled by the survey.  Of these 13
sources detected which lie outside the fully sampled parameter space
of the survey, 6 were new discoveries and 7 were already known.  Most
of the sources in the region $l$ = 330\degr -- 335\degr\/ which lie outside
the complete sample do so because their velocities are more negative
than -86~\kms\/ (our observed limit of velocity range where spectra
are centred at -30~\kms\/).

\begin{figure*}
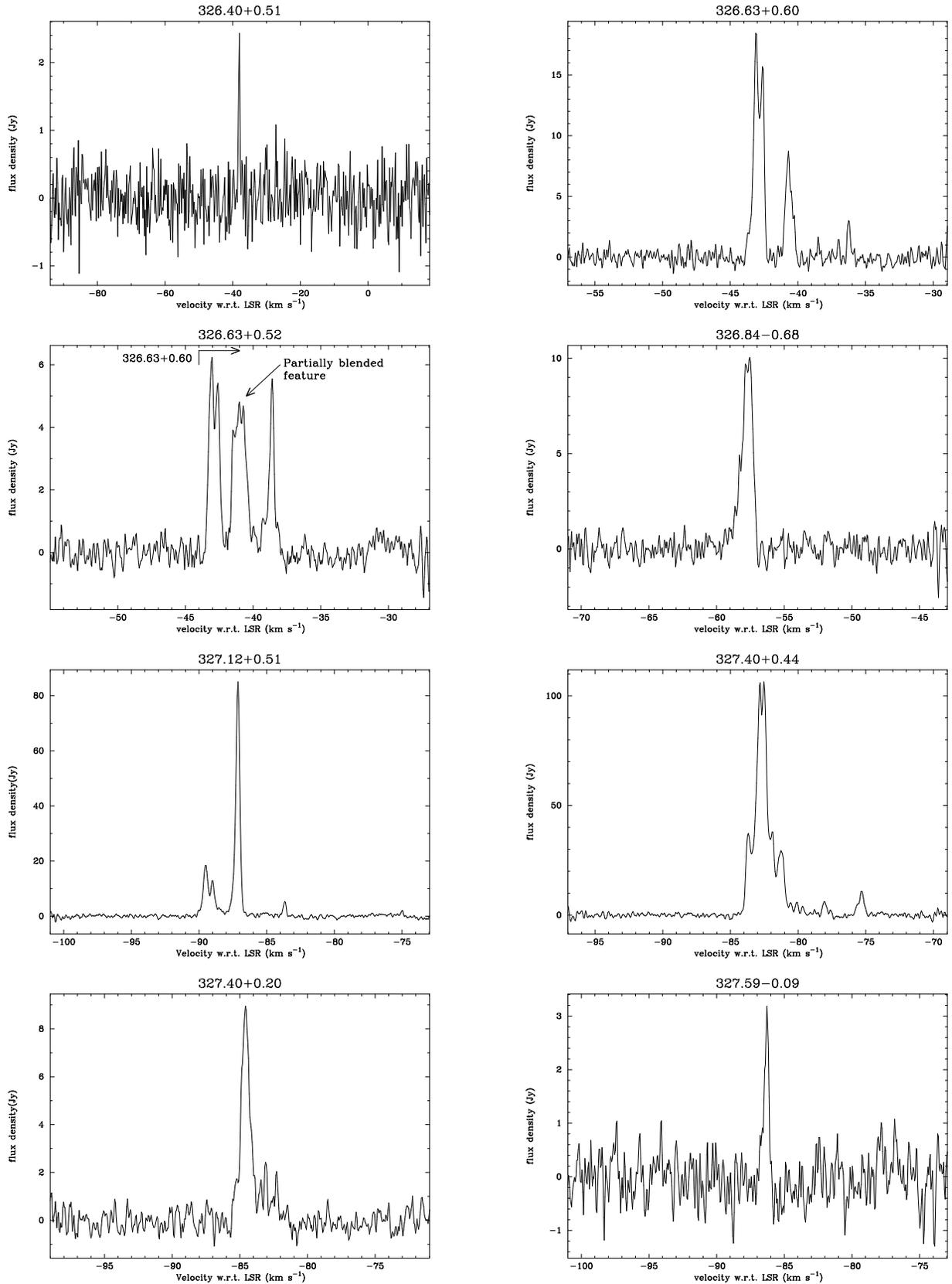

  \dfig{326_40}{326_63+0_60}
  \dfig{326_63}{326_84}
  \dfig{327_12}{327_40+0_44}
  \dfig{327_40}{327_59}
  \caption{Spectra of 6.7-GHz \methanol\/ masers detected in the region
	   $l$ = 325\degr -- 335\degr, $b$ = -0\fdg53 -- 0\fdg53}
  \label{fig:spectra}
\end{figure*}

\begin{figure*}
  \dfig{327_61-0_11}{327_93}
  \dfig{328_24}{328_25}
  \dfig{328_81}{329_03}
  \dfig{329_03-0_21}{329_07}
  \contcaption{}
\end{figure*}

\begin{figure*}
  \dfig{329_18}{329_33}
  \dfig{329_41}{329_48}
  \dfig{329_62}{330_95}
  \dfig{331_13}{331_28}
  \contcaption{}
\end{figure*}

\begin{figure*}
  \dfig{331_34}{331_42}
  \dfig{331_45}{331_54}
  \dfig{331_56}{332_11}
  \dfig{332_31}{332_33}
  \contcaption{}
\end{figure*}

\begin{figure*}
  \dfig{332_58}{332_95}
  \dfig{332_96}{333_03}
  \dfig{333_07}{333_12}
  \dfig{333_13}{333_15}
  \contcaption{}
\end{figure*}

\begin{figure*}
  \dfig{333_16}{333_20}
  \dfig{333_23}{333_33}
  \dfig{333_47}{333_58}
  \dfig{333_69}{333_95}
  \contcaption{}
\end{figure*}

\begin{figure*}
  \dfig{334_65}{335_08}
  \contcaption{}
\end{figure*}

Fig. 2 shows the positions of all the detected 6.7-GHz \methanol\/
masers, superimposed upon a 5-GHz continuum image of the region
\cite{Ha1978}.  In many cases the maser emission is closely associated
with local maxima in the continuum emission.  Sometimes the position
of the masers is slightly offset from the local maximum, probably at
the location of an ultra-compact core embedded in a cloud of more
diffuse gas.  Several centres of maser emission are often contained
within one \ionhy\/ region, indicating several separated sites of star
formation within the one larger molecular cloud.  In many cases, the
maser emission appears to be associated with the continuum emission.
Two regions in particular , 325\degr\/ $\lta l \lta$ 326\degr\/ and
329\fdg6 $\lta l \lta$ 331\fdg0, have a generally lower level of
continuum emission than the rest of the surveyed region and no masers
were detected in either region.  While not all of the continuum
features within the survey region have 6.7-GHz \methanol\/ masers
associated with them, approximately half do, and high-latitude
searches towards regions of enhanced continuum emission may prove to
be an effective searching method.  A comparison with the {\em IRAS}
Point-Source Catalog \shortcite{IRAS}, reveals that only 26 of the 50
sources detected have positions within 1~arcmin of an {\em IRAS}
source, and in many cases the nearest {\em IRAS} source is more than
2~arcmin away.

\begin{table*}
  \caption{6.7-GHz \methanol\/ masers in the region $l$=325\degr --
           335\degr\/, $b$=-0\fdg53 -- 0\fdg53.  References : 
             *=new source; 
             1=Batchelor \etal\/ (1980); 
             2=Caswell \etal\/ (1980); 
             3=Caswell \etal\/ (1993); 
             4=Caswell \etal\/ (1995a); 
             5=Caswell \etal\/ (1995c); 
             6=Peng \& Whiteoak (1992); 
             7=MacLeod \etal\/ (1992); 
             8=MacLeod \& Gaylard (1992); 
             9=Gaylard \& MacLeod (1993); 
             10=MacLeod, \etal\/ (1993); 
             11=Norris \etal\/ (1987); 
             12=Norris \etal\/ (1988);
             13=Norris \etal\/ (1993); 
             14=Schutte \etal\/ (1993);
             15=Slysh \etal\/ (1994); 
             16=Smits (1994)}
  \begin{tabular}{cccrrcrl}
  {\bf  Methanol}   & {\bf  Right Ascension}    & {\bf Declination} & 
    {\bf Peak} & {\bf Peak Vel.} & {\bf Velocity}  & {\bf Integrated} &
    {\bf references} \\
  {\bf maser}       & {\bf (J2000)}             & {\bf (J2000)}     & 
    {\bf Flux} & {\bf wrt LSR}   & {\bf Range}     & {\bf Flux~~~~} & \\
  {\bf ($l,b$)}     &                           &                   & 
    {\bf (Jy)} & {\bf (\kms\/)}  & {\bf (\kms\/)}  & {\bf (Jy\kms\/)} & \\ [2mm]
  \hline\hline
  $326.40\!+\!0.51$ & 15:43:40.7      & -54:19:28   &  2.4 & -38.1     &
	      &      1.4~~~~~~ & * \\
  $326.63\!+\!0.60$ & 15:44:33.3      & -54:06:23   &   18 & -43.1     & 
     -44,-36  &     18.8~~~~~~ & 1,14 \\
  $326.63\!+\!0.52$ & 15:44:52.1      & -54:10:30   &  5.6 & -41.0     &
     -42,-38  &      8.3~~~~~~ & *,6 \\
  $326.84\!-\!0.68$ & 15:51:11.1      & -54:59:01   &   10 & -57.6     & 
     -59,-57  &      9.7~~~~~~ & * \\
  $327.12\!+\!0.51$ & 15:47:33.6      & -53:52:35   &   80 & -87.1     & 
     -90,-83  &     49.8~~~~~~ & 1,2,4,8 \\
  $327.40\!+\!0.44$ & 15:49:14.0      & -53:45:36   &  106 & -82.6     & 
     -84,-75  &    153.1~~~~~~ & 1,2,3,4,7,11 \\
  $327.40\!+\!0.20$ & 15:50:21.4      & -53:56:25   &    9 & -84.6     & 
     -86,-82  &      9.0~~~~~~ & * \\
  $327.59\!-\!0.09$ & 15:52:34.0      & -54:03:12   &  3.2 & -86.2     & 
	      &      1.7~~~~~~ & * \\
  $327.61\!-\!0.11$ & 15:52:47.3      & -54:03:15   &  2.3 & -97.5     & 
	      &      1.8~~~~~~ & * \\
  $327.93\!-\!0.14$ & 15:54:34.8      & -53:52:18   &  6.6 & -51.7     & 
     -52,-51  &      3.0~~~~~~ & * \\
  $328.24\!-\!0.55$ & 15:57:58.5      & -53:59:23   &  421 & -44.9     & 
     -46,-34  &    437.7~~~~~~ & 1,3,4,6,7,11,13,15 \\
  $328.25\!-\!0.53$ & 15:57:59.9      & -53:58:01   &  425 & -37.4     & 
     -50,-36  &    428.1~~~~~~ & 1,3,4,6,7,11,13 \\
  $328.81\!+\!0.63$ & 15:55:51.2      & -52:42:36   &  278 & -44.5     & 
     -47,-43  &    353.0~~~~~~ & 1,3,4,7,11,13,15 \\
  $329.03\!-\!0.20$ & 16:00:22.1      & -53:12:57   &   25 & -41.9     & 
     -47,-41  &     38.2~~~~~~ & 1,3,4,7,11,15,16 \\
  $329.03\!-\!0.21$ & 16:00:36.0      & -53:12:17   &  275 & -37.5     & 
     -41,-34  &    360.0~~~~~~ & 1,3,4,7,11 \\
  $329.07\!-\!0.31$ & 16:01:12.7      & -53:15:58   &   24 & -43.9     & 
     -48,-43  &     17.4~~~~~~ & * \\
  $329.18\!-\!0.31$ & 16:01:36.3      & -53:11:49   &   13 & -55.7     & 
     -60,-51  &     18.6~~~~~~ & 1,2,4,8 \\
  $329.33\!+\!0.15$ & 16:00:29.3      & -52:44:39   &   14 & -106.5    & 
     -107,-10 &      8.3~~~~~~ & * \\
  $329.41\!-\!0.46$ & 16:03:36.5      & -53:08:58   &  144 & -66.8     & 
     -71,-66  &    106.0~~~~~~ & 1,2,4,8,10 \\
  $329.48\!+\!0.51$ & 15:59:39.9      & -52:22:45   &   13 & -72.1     & 
     -73,-65  &     17.1~~~~~~ & 14 \\
  $329.62\!+\!0.11$ & 16:02:07.3      & -52:35:13   &   30 & -60.1     & 
     -69,-59  &     26.5~~~~~~ & * \\
  $330.95\!-\!0.18$ & 16:09:53.3      & -51:55:38   &    7 & -87.6     & 
     -89,-87  &      4.1~~~~~~ & 1,2,4,6,9 \\
  $331.13\!-\!0.24$ & 16:11:00.7      & -51:51:18   &   34 & -84.4     & 
     -92,-84  &     43.3~~~~~~ & 1,2,4,5,8,15 \\
  $331.28\!-\!0.19$ & 16:11:22.3      & -51:42:26   &  165 & -78.1     & 
     -86,-78  &    250.3~~~~~~ & 2,3,4,5,7,11,12,13 \\
  $331.34\!-\!0.35$ & 16:12:23.3      & -51:46:11   &   66 & -67.4     & 
     -68,-64  &    124.9~~~~~~ & 2,4,8,15 \\
  $331.42\!+\!0.26$ & 16:10:10.3      & -51:16:18   &   25 & -88.6     & 
     -91,-88  &     19.9~~~~~~ & * \\
  $331.45\!-\!0.18$ & 16:12:14.6      & -51:34:39   &   70 & -88.5     & 
     -93,-84  &    158.0~~~~~~ & * \\
  $331.54\!-\!0.07$ & 16:12:10.9      & -51:25:24   &   12 & -84.1     & 
     -87,-83  &     12.5~~~~~~ & 1,2,4,6,7 \\
  $331.56\!-\!0.12$ & 16:12:28.7      & -51:27:04   &   35 & -103.4    & 
     -105,-94 &     46.0~~~~~~ & 4 \\
  $332.11\!-\!0.42$ & 16:16:19.3      & -51:17:50   &   16 & -61.4     & 
     -62,-58  &     16.8~~~~~~ & * \\
  $332.31\!-\!0.10$ & 16:15:50.5      & -50:55:26   &  6.3 & -47.0     & 
     -47,-42  &      8.3~~~~~~ & * \\
  $332.33\!-\!0.44$ & 16:17:29.3      & -51:09:24   &    4 & -53.8     & 
	      &      4.5~~~~~~ & * \\
  $332.58\!-\!0.15$ & 16:17:19.2      & -50:46:33   &    5 & -51.0     & 
     -56,-49  &      5.6~~~~~~ & * \\
  $332.95\!-\!0.68$ & 16:21:19.9      & -50:53:40   &   21 & -52.9     & 
     -54,-52  &     14.9~~~~~~ & * \\
  $332.96\!-\!0.68$ & 16:21:23.2      & -50:53:02   &   41 & -45.9     & 
     -48,-38  &     49.2~~~~~~ & * \\
  $333.03\!-\!0.02$ & 16:18:47.3      & -50:21:56   &    3 & -53.6     & 
     -61,-53  &      6.3~~~~~~ & * \\
  $333.07\!-\!0.45$ & 16:20:53.7      & -50:37:32   &   12 & -54.5     & 
     -55,-53  &      7.7~~~~~~ & 4,6 \\
  $333.12\!-\!0.43$ & 16:20:59.5      & -50:35:47   &   11 & -49.3     & 
     -50,-48  &     11.6~~~~~~ & 1,2,4,6,8,15 \\
  $333.13\!-\!0.44$ & 16:21:02.7      & -50:35:54   &    3 & -44.4     & 
     -45,-42  &      1.9~~~~~~ & 4,6 \\
  $333.15\!-\!0.56$ & 16:21:39.1      & -50:40:06   &   17 & -56.8     & 
     -63,-52  &     14.9~~~~~~ & * \\
  $333.16\!-\!0.10$ & 16:19:41.7      & -50:19:53   &    8 & -95.3     & 
     -95,-91  &      3.7~~~~~~ & 4 \\
  $333.20\!-\!0.08$ & 16:19:46.3      & -50:17:20   &  6.5 & -82.0     & 
     -85,-81  &      5.0~~~~~~ & 4 \\
  $333.23\!-\!0.06$ & 16:19:51.0      & -50:15:18   &  4.7 & -81.9     & 
     -92,-81  &      3.0~~~~~~ & 1,2,4,15 \\
  $333.33\!+\!0.11$ & 16:19:31.7      & -50:04:03   &    9 & -43.7     & 
     -50,-41  &     24.7~~~~~~ & * \\
  $333.47\!-\!0.17$ & 16:21:18.1      & -50:09:47   &   41 & -42.4     & 
     -49,-37  &     39.2~~~~~~ & 1,4,7 \\
  $333.58\!-\!0.02$ & 16:21:13.7      & -49:58:49   &   39 & -35.9     & 
     -37,-34  &     58.0~~~~~~ & * \\
  $333.69\!-\!0.44$ & 16:23:32.6      & -50:12:05   &   19 & -5.2      & 
     -6,-4    &     13.2~~~~~~ & * \\
  $333.95\!-\!0.14$ & 16:23:19.5      & -49:48:14   &    7 & -36.8     & 
     -37,-36  &      4.3~~~~~~ & * \\
  $334.65\!-\!0.02$ & 16:25:51.1      & -49:13:07   &   61 & -30.1     & 
     -31,-27  &     53.5~~~~~~ & * \\
  $335.08\!-\!0.43$ & 16:29:28.9      & -49:11:36   &   15 & -47.0     & 
     -48,-39  &     16.5~~~~~~ & * \\
  \end{tabular}
  \label{tab:main}
\end{table*}

\subsection{Comments on individual sources}

{\em 326.40+0.51}: This source is the weakest detected in the survey
and is the only source for which we were unable to improve the
positional accuracy from the initial survey observations.  There is no
associated {\em IRAS} source, the nearest being 15395-5411 which is
more than 3~arcmin away and has colours which suggest that it is unlikely
to be an UC\ionhy\/ region.

{\em 326.63+0.60}: This maser was discovered by Schutte \etal\/
\shortcite{Sc1993} (hereafter referred to as SWGM), who report it to
have a peak flux density of 70~Jy.  We measured a peak flux density of
18~Jy, implying a variation of more than a factor of 4 over a period
of 15 months.  A comparison of our spectrum with that of SWGM shows
that the ratio of the various features has changed surprisingly
little, considering the magnitude of the absolute variations.
However, we detect a 2-Jy peak at -36~\kms\/ which is beneath the
noise level in their spectrum.  SWGM detected this maser by searching
toward the {\em IRAS} source 15408-5356, but we measure the maser
emission to be separated from the IRAS source by 1.5~arcmin.

Molecular emission towards this region was first detected by Kaufmann
\etal\/ \shortcite{Ka1976}, who discovered \water\/ maser emission at
a velocity of -41 \kms\/.  Broad 12.2-GHz \methanol\/ absorption,
14.5-GHz \formaldehyde\/ absorption, recombination line emission and
44.1-GHz \methanol\/ maser emission, have also been detected at nearby
positions, all with velocities between -38 and -45~\kms\/
\cite{Pe1992,Sl1994}.  Although we cannot be certain that all these
sources are positionally coincident, their velocities suggest that
they are, and the unusual combination of emission and absorption
sources, a highly variable 6.7-GHz maser, and no detected OH maser
implies that this would be an interesting source for further
high-resolution study.

{\em 326.63+0.52}: This maser is separated by nearly 5~arcmin from
326.63+0.60.  However, there is an overlap in the velocity ranges, and
the peak component is partially blended with the weaker component of
326.63+0.60.  Again there is no nearby {\em IRAS} source, though the
nearest source 15412-5359, (separated by 2.2~arcmin from the maser
emission) is identified as an \ionhy\/ region, or dark cloud.

{\em 327.12+0.51}: This source is typical of stronger class II
\methanol\/ masers, with associated OH and \water\/ masers covering a
similar velocity range \cite{Ca1980,Ba1980}.  Caswell \etal\/
\shortcite{Ca1995a} detected no variability over the period of their
observations.  Our observations show that the source has remained
relatively constant in the intervening period, with perhaps a slight
decrease in the component at -83.5~\kms.

{\em 327.40+0.44}: This source was first detected at the position of a
known 12.2-GHz \methanol\/ maser by MacLeod \etal\/
\shortcite{Ma1992a}.  There is an anticorrelation between the the
strongest and weaker features of the 6.7- and 12.2-GHz \methanol\/
emission and those of the associated OH and \water\/ masers
\cite{Ca1993,Ca1980,Ba1980}.  The 6.7-GHz emission covers a greater
velocity range than any of the other transitions, which suggests that
the \water\/ and OH masers are part of a blue-shifted outflow.

{\em 327.40+0.20}: A new detection, this source appears to be
associated with the {\em IRAS} source 15464-5348, which lies in region
III of the colour--colour diagram as defined by SWGM.  However, it
fails their 60- and 100-\micron\/ flux density criteria and so they
did not search the {\em IRAS} source.

{\em 327.59-0.09 \& 327.61-0.11}: These two new sources are separated
by 2~arcmin.  Each has only a single component and neither has an {\em
IRAS} counterpart.

{\em 327.93-0.14}: This new detection has a single asymmetric feature
which is probably a blend of two or more components.  The closest {\em
IRAS} source (15507-5345), is separated by 1.6~arcmin from the maser
emission and has colours atypical of an UC\ionhy\/ region.

{\em 328.24-0.55 \& 328.25-0.53}: These two masers are the strongest
sources detected in this survey.  They are also in one of the best
studied regions of molecular emission in the Southern Hemisphere,
exhibiting maser emission from the 6.7-, 12.2- and 44.1-GHz
transitions of \methanol\, as well as OH \cite{Ca1993,Sl1994,Ca1980}.
High-resolution images of the 6.7-GHz emission made with the ATCA by
Norris \etal\/ \shortcite{No1993} show that the spatial morphology of
the maser spots is approximately linear at both centres of emission.
The 44.1-GHz emission is centred on a velocity of -41 \kms\/
\cite{Sl1994}.  However, at this velocity there is no emission from
the other \methanol\/ transitions and the 12.2-GHz absorption is
strongest \cite{Ca1995c}.  This is consistent with current theories of
class I \methanol\/ maser emission which suggest that they are
partially collisionally pumped \cite{Cr1992}, possibly by
high-velocity outflows interacting with the parent molecular cloud
\cite{Me1993}.  These sources are listed as slightly variable at
6.7~GHz by Caswell \etal\/ \shortcite{Ca1995a}, and the peak flux
density we measure for 328.24-0.55 appears to be slightly larger than
that observed by Caswell \etal

{\em 328.81+0.63}: This source is also strong and well studied,
showing maser emission from the 6.7-, 12.2- and 44.1-GHz transitions of
\methanol\/ and the 1.665- and 6.035-GHz transitions of OH
\cite{Ca1993,Sl1994,Ca1980,Sm1994}.  It is a good example of the
radial velocity anticorrelation between class I and class II
\methanol\/ masers, noted by Slysh \etal\/, with the 6.7-GHz emission
covering the range -47 to -43~\kms\/ and the 44.1-GHz emission covering
the range -43 to -40~\kms\/.  The peak of the OH emission is at a
similar velocity to that of the class II \methanol\, but the total
velocity range is larger, encompassing the ranges of both the class I
and II \methanol\/ masers.  328.81+0.63 is another source for which
high-resolution studies show the spatial distribution of the 6.7-GHz
maser spots to be highly linear \cite{No1993}.  Most of the
maser species associated with this source are relatively strong,
making it an excellent candidate for studying the relationship
between the different molecular transitions, and star formation in
general.  Comparison of our spectrum with that of Caswell \etal\/
\shortcite{Ca1995a}, shows a decrease in the peak flux density of
approximately 25~per cent over the last two years.

{\em 329.03-0.20 \& 329.03-0.21}: Maser emission from a wide range of
transitions (12.2- and 44.1-GHz \methanol\/ and OH) is associated with
these two sources, though primarily with 329.03-0.21, which is the
stronger at 6.7~GHz \cite{No1987,Sl1994,Ca1980}.  The closest {\em
IRAS} source to both centres of maser emission is 15566-5304, which is
not likely to be an UC\ionhy\/ region, as it has a 12-\micron\/ flux
density which is greater than the 25-\micron\/ flux density.  ATCA
observations by S. P. Ellingsen \etal\/ (in preparation) failed to detect
any continuum emission associated with either of the masing regions.

{\em 329.07-0.31}: This source is close to 329.03-0.20 \& 329.03-0.21,
and a sidelobe response from the latter can been seen quite clearly in
the spectrum.  Although the velocity ranges of 329.03-0.20 and
329.07-0.31 overlap, 329.03 appears to be sufficiently distant that it
does not contribute significantly to the observed flux density of
329.07-0.31.  The maser appears to be associated with the {\em IRAS}
source 15573-5307, which falls in region I of SWGM's {\em IRAS}
colour--colour diagram.  However, the 60- and 100-\micron\/ flux
densities fail their criteria.

{\em 329.18-0.31}: This is another well-known maser source, with
associated 12.2-GHz \methanol\/, OH and \water\/ maser emission
\cite{Ca1993,Ca1980,Ba1980}.  It exhibits much weaker \methanol\/
maser emission in both class II transitions than most other sources in
this category.

{\em 329.33+0.15}: This new detection appears to be associated with
the {\em IRAS} source 15567-5236, with the 6.7-GHz \methanol\/ maser
situated 0.6~arcmin from the {\em IRAS} source.  This source is listed
as a non-detection by SWGM, which implies that in mid-1993 it had a
flux density of $<$ 5~Jy.  We observed a peak flux density of
14~Jy, which implies a lower limit on the increase of nearly 3 in
less than a year.

{\em 329.41-0.46}: This object shows OH and \water\/ maser emission, and
both absorption and weak emission from 12.2-GHz \methanol\/
\cite{Ca1980,Ba1980,Ca1995c}.  The strongest absorption is redshifted
with respect to the peak emission in all transitions, although there
is some OH emission in this region which may represent an outflow.

{\em 329.48+0.51}: This is the second maser in our sample discovered
by SWGM.  It has a somewhat unusual spectral appearance, with at least
six maser features within a 3~\kms\/ velocity range and one further
feature blueshifted by 5~\kms\/ from the other emission.  The offset
feature is not a separate source, as far as we can determine from our
single-dish grid observations.

{\em 330.95-0.18}: This source is unusual in that the OH and \water\/
maser emission is much more complex and covers a far greater velocity
range than that of the 6.7-GHz \methanol\/ \cite{Ca1980,Ba1980}.  It
is also one of the few sources for which the OH maser emission is
stronger than the 6.7-GHz \methanol.  Studying this source in detail
and comparing it with other more typical sources may yield important
insights into the physical conditions required to pump both OH and
\methanol\/ masers.

{\em 331.13-0.24}: This is one of the most variable 6.7-GHz
\methanol\/ masers, with some features changing by more than an order
of magnitude over a period of a few months \cite{Ca1995b}.  Comparison
of our spectrum with those published in the literature shows that
large-scale variations are continuing \cite{Ma1992b,Ca1995a,Ca1995b}.
The \water\/ maser emission is quite weak and spans a larger velocity
range than the \methanol\/, while the OH emission has its peak in
between the two centres of \methanol\/ emission \cite{Ba1980,Ca1980}.
The 44.1-GHz \methanol\/ maser also has several discrete regions of
emission, the peak corresponding to the the weaker of the two regions
in our 6.7-GHz spectra and the secondary approximately coincident with
the peak of the OH emission \cite{Sl1994}.

{\em 331.28-0.19}: This source is one of the better studied
\methanol\/ maser sources, with high-resolution 6.7- and 12.2-GHz
images showing a positional coincidence of some of the aligned
spectral features to better than 20~milliarcsec \cite{No1988,No1993}.
Caswell \etal\/ \shortcite{Ca1995a} list this source as slightly
variable, and a comparison of our spectrum with that of Caswell
\etal\/ shows that the strongest component appears to have decreased by
20~per cent over the last two years.  This source also exhibits
strong maser emission in the 12.2-GHz transition of \methanol\/.  As
the peak components at 6.7 and 12.2~GHz exhibit both a spectral and
positional coincidence, a comparison of the variability at the two
frequencies may yield important information on the pumping mechanism
for the class II \methanol\/ masers.

{\em 331.34-0.35}: Caswell \etal\/ \shortcite{Ca1995b} found that the
stronger 6.7-GHz \methanol\/ masers are typically less variable than
the weaker maser sources.  However, 331.34-0.35 appears to be an exception.  A
comparison of our spectrum with that of Caswell \etal\/
\shortcite{Ca1995a} shows that the feature at -65 \kms\/ has decreased
by about a factor of 2, while at the same time the feature at
-68 \kms\/ has increased by a factor of 2 and the feature at -67
\kms\/ has remained relatively unchanged.  Unusually, the class I 44.1-GHz
\methanol\/ maser emission lies in the same velocity range as the
class II 6.7-GHz emission \cite{Sl1994}.

\begin{figure*}
  \centering
  \begin{minipage}[t]{0.50\textwidth}
    \epsfxsize=0.99\textwidth
    \epsout 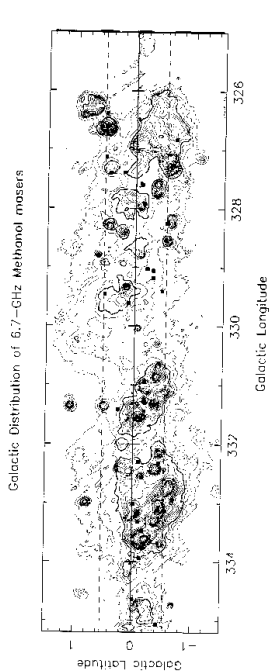
  \end{minipage}
  \caption{5-GHz continuum emission $l$ = 325\degr -- 335\degr\/, $b$ =
	   -1\fdg5 -- 1\fdg5 (Haynes \etal\/ 1978).  The positions of
	   the 6.7-GHz \methanol\/ masers are marked with squares and
	   the dashed line marks the latitude extent of the survey.
           {\bf NOTE:} A higher resolution, full page version of this figure
           can be obtained from 
           http://reber.phys.utas.edu.au/$\sim$sellings/preprints.html}
  \label{fig:cont}
\end{figure*}

{\em 331.42+0.26}: This new detection lies at the edge of a large
region of complex continuum emission (see Fig.~\ref{fig:cont}).  The
nearest {\em IRAS} source 16062-5108 is 1~arcmin away and has
colours atypical of an \ionhy\/ region.

{\em 331.45-0.18}: This is the strongest 6.7-GHz \methanol\/ maser of those
discovered by this survey and has a complex spectral
morphology.  It lies in the region of the Galactic Plane which Caswell
\etal\/ \shortcite{Ca1980} searched for OH maser emission, implying
that there was no OH maser with peak flux density $>$ 1~Jy.  Unless OH
emission is present but was missed in the earlier surveys, this source
lies at an extreme of the 6.7-GHz \methanol:OH flux ratio
distribution.  This source lies 0.6~arcmin from an {\em IRAS} source
(16084-5127) which has colours placing it in region III of SWGM's {\em
IRAS} colour--colour diagram, but fails their 100-\micron\/ flux
density criterion.

{\em 331.54-0.07 \& 331.56-0.12}: The first of these sources was
discovered by Caswell \etal\/ \shortcite{Ca1995a} and is near
331.56-0.12, which can be seen at the edge of our spectrum.  There is
OH and \water\/ maser emission in the general direction of these two
sources, though exactly which features are associated with each of the
\methanol\/ sources has not yet been clearly determined
\cite{Ca1995a}.  The OH emission has a velocity range from
approximately -85 to -94 \kms\/ and the \water\/ emission has a
velocity range from approximately -75 to -105 \kms\/ with a
blueshifted outflow at a velocity of approximately -140 \kms\/
\cite{Ca1980,Ba1980}.  331.56-0.12 also has a 12.2-GHz \methanol\/
maser associated with it which has a similar spectral morphology to
the 6.7-GHz maser \cite{Ca1995c}.

{\em 332.11-0.42}: This is the first of several new 6.7-GHz
\methanol\/ masers discovered between Galactic latitudes 332\degr\/
and 333\degr\/.  This source has a good positional coincidence with
the {\em IRAS} source 16124-5110, which has colours atypical of an
UC\ionhy\/ region.

{\em 332.31-0.10}: This new detection is close to the {\em IRAS}
source 16119-5048.  This meets the colour and flux density criteria
for region III of SWGM's {\em IRAS} colour--colour diagram, but it has
an upper limit for the 100-\micron\/ flux density and so was not
searched by them.

{\em 332.33-0.44}: This maser lies at the edge of a peak in the
continuum emission (see Fig.~\ref{fig:cont}), but does not appear to
have an {\em IRAS} counterpart.  The nearest {\em IRAS} source
(16357-5100) is 1.3~arcmin away and has colours which suggest
that it is probably not an UC\ionhy\/ region.

{\em 332.58-0.15}: This source also lies at the edge of a continuum
peak, but has no {\em IRAS} counterpart, the nearest {\em IRAS} source
16136-5038 being 2~arcmin away.

{\em 332.95-0.68 \& 332.96-0.68}: Discovered serendipitously while
taking a reference spectrum for another observation, these sources are
separated by only 48~arcsec, but have non-overlapping velocity ranges.
332.95-0.68 appears to be associated with the {\em IRAS} source
16175-5046, and 332.96-0.68 with 16175-5045, both of which lie in
region III of SWGM's {\em IRAS} colour--colour diagram.  However, each
has upper limits for two of the {\em IRAS} flux density bands and so
neither {\em IRAS} source was searched by them.

{\em 333.03-0.02}: This is a 6.7-GHz \methanol\/ weak maser which lies
near the edge of a continuum peak (see Fig.~\ref{fig:cont}), and has
no {\em IRAS} counterpart.  It differs from most sources of this
strength in that it has at least four components.

{\em 333.07-0.45, 333.12-0.43 \& 333.13-0.44}: Although positionally
adjacent, these sources are widely separated in velocity.  There is
12.2- and 44.1-GHz \methanol\/ maser emission as well as OH and \water\/
maser emission from this general region
\cite{Ca1995c,Sl1994,Ca1980,Ba1980}.  Caswell \etal\/
\shortcite{Ca1995c} detected weak 12.2-GHz emission associated with
333.07-0.54 and Caswell \etal\/ \shortcite{Ca1995a}, determined that
the OH emission is associated with 333.12-0.43, making this another
case where the OH emission is stronger that of the 6.7-GHz \methanol\/
emission and also covers a wider velocity range.  The 44.1-GHz
\methanol\/ and \water\/ maser emission cover the velocity ranges 
of both 333.12-0.43 and 333.13-0.44, and may not be closely associated
with either source.

{\em 333.15-0.56}: This new detection shows a somewhat unusual
spectral morphology, with three distinct single peaks.  Its position
is close to a peak in the continuum emission (see
Fig.~\ref{fig:cont}), but has no associated {\em IRAS} source.

{\em 333.16-0.10, 333.20-0.08 \& 333.23-0.06}: These sources are close
together and the latter two are severely blended in the 7-arcmin beam
of the Hobart telescope.  Caswell \etal\/ \shortcite{Ca1995a} list
all of the components of 333.23-0.06 as being variable and they
measure a flux density of 3.8~Jy for the -80.5~\kms\/ peak, whereas in
our spectrum it has a flux density of $<$ 1~Jy.  The strongest
peak in 333.23-0.06 is at a velocity of -81.9~\kms\/ but it is
blended with emission from 333.20-0.08, and the true flux density of
this component is probably less than half that listed in
Table~\ref{tab:main}.  Once again there is emission from 12.2- and
44.1-GHz \methanol\/, OH and \water\/
\cite{Ca1995c,Sl1994,Ca1980,Ba1980}.  Caswell \etal\/
\shortcite{Ca1995a} determined that the OH emission is most closely
associated with 333.23-0.06 and this also appears to be the case for
the \water\/ and 44.1-GHz \methanol\/ maser emission.  This is somewhat
surprising as it is the weakest of this group of sources at 6.7~GHz.

{\em 333.33+0.11}: An unusually large number of components
distinguishes this source from most others with low peak flux density.
We made several observations of this source while trying to determine
its position and found its flux density to vary significantly over a
period of a week.  It appears to be associated with the {\em IRAS}
source 16157-4957, which lies in region I of SWGM's colour--colour
diagram, but does not satisfy their 60-\micron\/ flux density
criterion.

{\em 333.47-0.17}: This maser has a weak OH counterpart, with emission
and absorption spanning the range of most of the 6.7-GHz \methanol\/
emission \cite{Ca1980}.  Comparison of our spectrum with that of
Caswell \etal\/ \shortcite{Ca1995a} shows that the flux density of
the peak component in this source has nearly halved in intensity (from
70 to 41 Jy) over a period of 18 months, but the rest of the features
have remained relatively constant.  This source appears to be
associated with the {\em IRAS} source 16175-5002, which is 0.4~arcmin
away and has colours which place it in region III of SWGM's
colour--colour diagram.

{\em 333.58-0.02}: This is one of the stronger new sources, showing
several strong peaks within a small velocity range.  There are also
several weak features blueshifted from the peak emission by a few
\kms\/.  This source has no {\em IRAS} counterpart.

{\em 333.69-0.44}: This sources is associated with a peak in the
continuum emission (see Fig.~\ref{fig:cont}) and with the {\em IRAS}
source 16196-5005.  The {\em IRAS} source lies in region III of SWGM's
colour--colour diagram, but has upper limits for the 25- and
100-\micron\/ flux density measurements and so was not included in
their search.  This source has a velocity quite close to the local
standard of rest and the {\em IRAS} counterpart is identified as an
\ionhy\/ region or dark cloud.

{\em 333.95-0.14}: This new 6.7-GHz \methanol\/ maser appears to be
associated with the {\em IRAS} source 16194-4941, which lies well
outside the region of the colour--colour diagram which SWGM searched.
Like the previous source, the {\em IRAS} counterpart is identified as
an \ionhy\/ region, or a dark cloud.

{\em 334.65-0.02}: One of the strongest new sources detected, it may
be associated with the {\em IRAS} source 16220-4906, which does not
have the colours of an UC\ionhy\/ region, as the 25-\micron\/ flux
density is less than the 12-\micron\/ flux.

{\em 335.08-0.43}: This source lies at the edge of a large area of
low-level continuum emission (see Fig.~\ref{fig:cont}).  It does not
appear to be associated with an {\em IRAS} source, the closest being
16256-4905.

\section{Discussion}

This survey has doubled the number of 6.7-GHz \methanol\/ masers
detected in the region $l$ = 325\degr -- 335\degr\/, $b$ = -0\fdg53 --
0\fdg53.  Before this survey there were nearly 250 known 6.7-GHz
\methanol\/ masers in the Galaxy.  If we extrapolate the results of
the region we have surveyed to the entire Galaxy, this implies that
there are at least 500 detectable masers of the \transa\/ transition
of \methanol.  

Previous searches for 6.7-GHz \methanol\/ masers have not detected any
which are known to be associated with objects other than sites of
massive star formation, presumably because of the selection criteria
in previous searches.  Many of the detected sources are not associated
with {\em IRAS} or radio continuum sources, and so we have no
information on whether these are associated with star formation
regions or with some other type of object.  We used the Simbad
database to search for all sources within a 3-arcmin radius of the
6.7-GHz \methanol\/ maser detected in this survey, but found no
convincing associations with other classes of object.  To confirm this
requires observations toward the new maser detections in other regions
of the electromagnetic spectrum, including high-resolution radio
images and searches for other maser transitions.

\subsection{Associations with {\em IRAS} sources}

This survey detected 50 6.7-GHz \methanol\/ masers, of which 26 have
an {\em IRAS} source within 1~arcmin.  The details of these {\em IRAS}
sources are summarized in Table~\ref{tab:iras}. The following
discussion is confined to only those sources in the region $l$ =
325\degr -- 335\degr\/, $b$ = -0\fdg53 -- 0\fdg53 (the ``survey
region'') which we have surveyed completely.  However, we include
those sources in this region which have velocities outside the 
range completely surveyed.  Where we have calculated the fraction of
{\em IRAS} sources with an associated 6.7-GHz \methanol\/ maser, that
figure applies to searches with a sensitivity limit comparable to this
survey and is a lower limit for more sensitive searches.

To assess the various {\em IRAS}-based selection methods more
rigorously, we chose to compare the {\em IRAS} sources associated with
6.7-GHz \methanol\/ masers with all {\em IRAS} sources contained
within the survey region.  We detected 41 6.7-GHz \methanol\/ masers
within the survey region, of which 21 are within 1~arcmin of an {\em
IRAS} source.  There are a further 11 6.7-GHz \methanol\/ masers
separated by between 1 and 2~arcmin from an {\em IRAS} source.  Of
these 11 {\em IRAS} sources three also have a maser within 1~arcmin.
Many of these {\em IRAS} sources have colours typical of ultra-compact
\ionhy\/ regions and a search toward the {\em IRAS} position with the
7-arcmin beam of the Hobart telescope would have detected a maser
source.  For the purposes of our analysis we will assume that all
maser sources 2~arcmin or more from an {\em IRAS} source are
unassociated.  For the ranges listed below, the lower limit has been
calculated using only the maser sources within 1~arcmin, and the upper
limit with all sources $<$~2~arcmin from an {\em IRAS} source.  A
search of the {\em IRAS} Point-Source Catalog \shortcite{IRAS} found
876 sources contained in the survey region.  If we assume a uniform
distribution of {\em IRAS} sources within the survey region, then
there is 7~per cent probability of any 6.7-GHz \methanol\/ masers
being within 1~arcmin of an {\em IRAS} source.  Thus, for our sample
of 41 6.7-GHz \methanol\/ masers we would expect three chance
associations with {\em IRAS} sources.

\begin{table*}
  \begin{centering} 
  \caption{{\em IRAS} sources associated with 6.7-GHz \methanol\/
	   masers.  Those which fall outside the completely surveyed
	   region are marked with an asterisk.}
  \label{tab:iras}
  \end{centering}
  \begin{tabular}{ccrrrrrr}
  {\bf Methanol} & {\bf {\em IRAS}} &                    & {\bf Flux}       &
		     &                   &                               &
				  \\
  {\bf Maser}    & {\bf Name}       & {\bf 12~\micron}   & {\bf 25~\micron} & 
    {\bf 60~\micron} & {\bf 100~\micron} & $\bf Log_{10}(S_{25}/S_{12})$ & 
    $\bf Log_{10}(S_{60}/S_{12})$ \\
  ($\bf l,b$)    &                  & {\bf (Jy)}         & {\bf (Jy)} & 
    {\bf (Jy)}       & {\bf (Jy)}        &                               & 
				  \\ [2mm]
  \hline\hline
  $327.12\!+\!0.51\!$         & 15437-5343 &   6 &   75 &   988 &  1425 &  1.10 & 2.22 \\
  $327.40\!+\!0.44\!$         & 15454-5335 &   4 &   56 &  1153 &  2697 &  1.16 & 2.47 \\
  $327.40\!+\!0.20\!$         & 15464-5348 &   2 &   10 &   109 &   397 &  0.77 & 1.83 \\
  $328.25\!-\!0.53^{*}\!$     & 15541-5349 &  12 &  111 &  3033 &  6415 &  0.96 & 2.40 \\
  $328.81\!+\!0.63^{*}\!$     & 15520-5234 &  16 &  538 & 10780 & 16380 &  1.54 & 2.84 \\
  $329.03\!-\!0.21\!$         & 15566-5304 &   6 &    4 &   332 &  1652 & -0.19 & 1.72 \\
  $329.07\!-\!0.31\!$         & 15573-5307 &   3 &   17 &   124 &  1145 &  0.74 & 1.60 \\
  $329.33\!+\!0.15\!$         & 15567-5236 & 196 & 1077 &  7398 &  8360 &  0.74 & 1.58 \\
  $329.41\!-\!0.46\!$         & 15596-5301 &   5 &   52 &  1102 &  2487 &  1.03 & 2.36 \\
  $331.28\!-\!0.19\!$         & 16076-5134 &  36 &  237 &  2823 &  5930 &  0.82 & 1.89 \\
  $331.34\!-\!0.35\!$         & 16085-5138 &  41 &  285 &  2262 &  4841 &  0.84 & 1.74 \\
  $331.42\!+\!0.26\!$         & 16062-5108 &   3 &    2 &    22 &   240 & -0.08 & 0.95 \\
  $331.45\!-\!0.18\!$         & 16084-5127 &   2 &    9 &   179 &   225 &  0.72 & 2.02 \\
  $331.56\!-\!0.12\!$         & 16086-5119 &  15 &  161 &  1229 & 25800 &  1.03 & 1.91 \\
  $332.11\!-\!0.42\!$         & 16124-5110 & 119 &  290 &  6234 &  8651 &  0.39 & 1.72 \\
  $332.31\!-\!0.10\!$         & 16119-5048 &  11 &  108 &   926 &  2605 &  1.01 & 1.94 \\
  $332.95\!-\!0.68^{*}\!$     & 16175-5046 &   6 &   26 &   539 &  2194 &  0.68 & 1.99 \\
  $332.96\!-\!0.68^{*}\!$     & 16175-5045 &   7 &   31 &   562 &  1820 &  0.65 & 1.90 \\
  $333.12\!-\!0.43\!$         & 16172-5028 & 144 & 1514 & 12380 & 26700 &  1.02 & 1.93 \\
  $333.13\!-\!0.44\!$         & 16172-5028 & 144 & 1514 & 12380 & 26700 &  1.02 & 1.93 \\
  $333.16\!-\!0.10\!$         & 16159-5012 &   4 &   41 &   817 &  3569 &  0.98 & 2.28 \\
  $333.33\!+\!0.11\!$         & 16157-4957 &   5 &   39 &   256 &  3691 &  0.94 & 1.76 \\
  $333.47\!-\!0.17\!$         & 16175-5002 &  11 &   77 &  1135 &  3210 &  0.87 & 2.03 \\
  $333.69\!-\!0.44\!$         & 16196-5005 &   3 &   22 &   306 &  1120 &  0.81 & 1.95 \\
  $333.95\!-\!0.14\!$         & 16194-4941 &   9 &   20 &   207 &   555 &  0.33 & 1.35 \\
  $334.65\!-\!0.02\!$         & 16220-4906 &   4 &    3 &    76 &   304 & -0.09 & 1.31 \\
  \end{tabular}
\end{table*}

\begin{figure*}
  \centering
  \begin{minipage}[t]{0.80\textwidth}
    \epsfxsize=0.99\textwidth
    \epsout 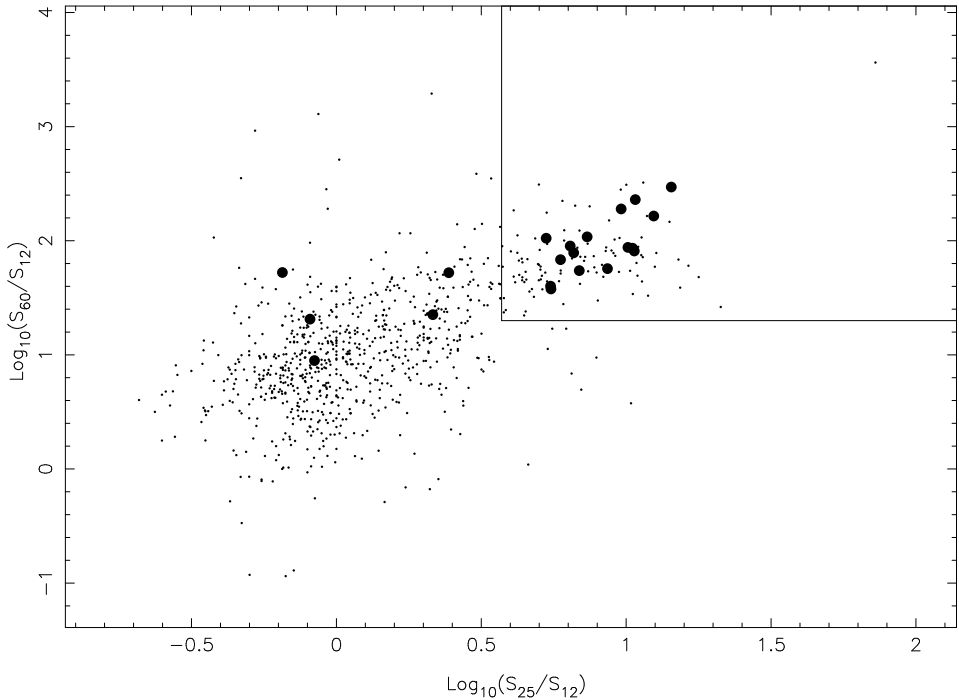
  \end{minipage}
  \caption{Distribution of 60/12 versus 25/12~\micron\/ colours for all
	   {\em IRAS} sources in the region $l$ = 325\degr -- 335\degr\/,
	   $b$ = -0\fdg53 -- 0\fdg53.  The sources with a 6.7-GHz
	   \methanol\/ maser within 1~arcmin are marked with a filled
	   circle.}
  \label{fig:iras1}
\end{figure*}

Fig.~\ref{fig:iras1} shows a plot of 60/12~\micron\/ versus
25/12~\micron\/ colours for all 876 {\em IRAS} sources in our survey
region.  Those associated with 6.7-GHz \methanol\/ masers are marked
with a filled circle.  Sixteen of the sources associated with masers
lie in the upper right corner of this colour--colour diagram.  Wood \&
Churchwell \shortcite{Wo1989} have shown that there is a high
probability that the sources in this area of the colour--colour
diagram are ultra-compact \ionhy\/ regions (UC\ionhy\/ regions).  In
total there are 120 {\em IRAS} sources in our sample which satisfy the
criteria of $Log_{10}(S_{60}/S_{12}) \geq
1.30$~\&~$Log_{10}(S_{25}/S_{12}) \geq 0.57$ (the solid lines in
Fig.~\ref{fig:iras1}).  We will call these criteria the Wood \&
Churchwell minus (WC-) criteria (as they are less stringent than the
full Wood \& Churchwell criteria).  For each flux density measurement
the {\em IRAS} catalog contains a quality flag in each wavelength
band.  Having selected sources according to the WC- criteria, Wood \&
Churchwell then excluded those for which either the 25- or
60-\micron\/ flux density measurement is only an upper limit, as they
are most likely to be situated lower and further left in the
colour--colour diagram than their current position.  We call these
criteria the WC criteria.

Assuming that the {\em IRAS} sources selected using WC- are uniformly
distributed throughout the survey region, the probability of detecting
a 6.7-GHz \methanol\/ maser within 1~arcmin is $\sim$ 1~per cent.
Thus, for our sample of 41 6.7-GHz \methanol\/ masers, we would
expect 0.4 chance associations with an {\em IRAS} source in the Wood
\& Churchwell region of the colour--colour diagram.  Of the 120
sources which satisfy the WC- criteria, 16 are within 1~arcmin and 19
within 2~arcmin of a 6.7-GHz \methanol\/ maser.  If we assume that
there are no chance associations, then this implies that any search
based on these criteria will detect 6.7-GHz \methanol\/ masers
associated with 13--16~per cent of the selected {\em IRAS} sources.

If we apply the WC criteria, we are left with 74 sources, 14 of which
are within 1~arcmin and 17 within 2~arcmin of a 6.7-GHz \methanol\/
maser source.  If we again assume no chance associations, this implies
that 19--23~per cent of the {\em IRAS} sources selected using these
criteria will have an associated 6.7-GHz \methanol\/ maser.

The only published search for 6.7-GHz \methanol\/ masers toward {\em
IRAS} sources, used selection criteria based on that of Wood \&
Churchwell \shortcite{Wo1989}, but with varying lower limits on the
60- and 100-\micron\/ fluxes (SWGM).  If we apply their selection
criteria to our sample we find 31 sources which satisfy their
criteria, 10 of which are within 1~arcmin and 13 within 2~arcmin of a
6.7-GHz \methanol\/ maser.  Thus, for a sample of {\em IRAS} sources
selected using SWGM's criteria, we would expect 32--42~per cent to
have an associated 6.7--GHz \methanol\/ maser.  The other sources
which lie inside the Wood \& Churchwell UC\ionhy\/ region of the
colour--colour diagram fail one or more of the flux density or flux
quality criteria.  The distribution of the Galactic latitudes at which
SWGM detected new 6.7-GHz \methanol\/ masers is relatively flat for
$|b|$ \lta 1.0\degr.  This implies that {\em IRAS} colour-based
selection criteria may be the most practical for finding sources which
are not close to the Galactic Plane.

Hughes \& MacLeod \shortcite{Hu1989}, developed an independent method
of identifying \ionhy\/ regions on the basis of their {\em IRAS}
colours, which they claim has a confidence level of 89~per cent.  Their
criteria were that $Log_{10}(S_{25}/S_{12}) \geq
0.40$,~$Log_{10}(S_{60}/S_{25}) \geq 0.25$,~$S_{100} \geq 80$~Jy and
that the flux quality flag for the 25-, 60- and 100-\micron\/ bands
was not an upper limit.  We call these criteria the HM criteria.  Of
the 876 {\em IRAS} sources in the ``survey region'', 69 meet the HM
criteria ; 11 of these 69 {\em IRAS} sources are within 1~arcmin and
15 within 2~arcmin of a 6.7-GHz \methanol\/ maser. If we assume that
there are no chance associations, this implies that 16--22~per cent of
the {\em IRAS} sources selected using the HM criteria will have an
associated 6.7-GHz \methanol\/ maser.

756 {\em IRAS} sources are outside the UC\ionhy\/ region of the
colour--colour diagram. Of these 5 are within 1~arcmin of a 6.7-GHz
\methanol\/ maser (compared to the 3 expected by chance alone).
Therefore only 1~per cent of {\em IRAS} sources in this region of the
colour--colour diagram have an associated 6.7-GHz \methanol\/ maser.

On the basis of their {\em IRAS} colours Palla \etal\/
\shortcite{Pa1991} conducted a search for \water\/ masers toward a
selection of {\em IRAS} sources, identified as candidate UC\ionhy\/
regions or dense molecular clouds.  They separated their detections
into two groups, those which satisfied the WC criteria and those which
did not. This latter group they labeled ``low''.  Although the 5 {\em
IRAS} sources with associated 6.7-GHz \methanol\/ masers which fail
the WC criteria occupy a similar region of the colour--colour diagram
to the ``low'' sample of Palla \etal\/ \shortcite{Pa1991}, they differ
in other ways.  In particular, Palla \etal\/ found that the masers
associated with {\em IRAS} sources outside the Wood \& Churchwell
region of the colour--colour diagram were in general weaker than those
inside.  In contrast, we find that both the median and mean of the
peak flux densities is greater for those sources outside the Wood \&
Churchwell region than those inside.  However, the outside sample is
small and the mean in particular is dominated by the flux density of
329.03-0.21. We have shown above that we expect 3 chance associations
of 6.7-GHz \methanol\/ masers with {\em IRAS} sources within 1~arcmin,
and there is a 15.6~per cent probability of 5 or more chance
associations from 41 sources.  We therefore consider it likely that
these associations are due to chance, although further observations
are required to confirm this.

\begin{figure*}
  \centering 
  \begin{minipage}[t]{0.80\textwidth}
    \epsfxsize=0.99\textwidth
    \epsout 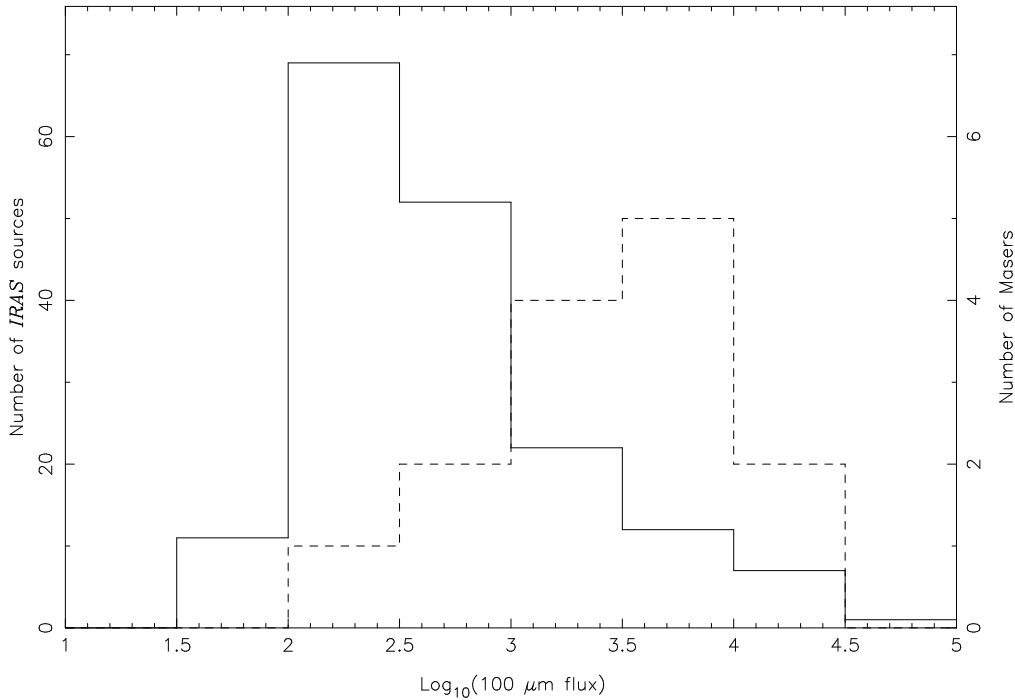
  \end{minipage}
  \caption{The solid line shows the distribution of 100-\micron\/ flux
	   for {\em IRAS} sources in the region $l$ = 325\degr --
	   335\degr\/, $b$ = -0\fdg53 -- 0\fdg53, with a moderate or
	   higher flux quality.  The dashed-line shows the
	   distribution of the sources with 6.7-GHz \methanol\/ masers
	   within 1~arcmin, and the vertical scale multiplied by 10.}
  \label{fig:iras2}
\end{figure*}

Fig.~\ref{fig:iras2} and Table~ \ref{tab:100micronhist} compare the
distribution of 100-\micron\/ flux density (excluding those sources
which have only an upper limit for the 100-\micron\/ flux density) of
the {\em IRAS} sources with associated 6.7-GHz \methanol\/ masers,
with that of the {\em IRAS} sources in the region.  The probability of
maser association clearly increases rapidly with increasing
100-\micron\/ flux.  In the ``survey region'' 132 {\em IRAS} sources
have a 100-\micron\/ flux density of $<$ 1000~Jy, of which 3 are
within 1~arcmin and 4 within 2~arcmin of a 6.7-GHz \methanol\/ maser.
By comparison, of the 42 {\em IRAS} sources with a 100-\micron\/ flux
density $>$ 1000~Jy, 11 are within 1~arcmin and 14 within 2~arcmin of
a 6.7-GHz \methanol\/ maser.  This implies that the probability of
detecting a 6.7-GHz \methanol\/ maser associated with an {\em IRAS}
source is $\sim$ 2~per cent if the source has a 100-\micron\/ flux
density $<$ 1000~Jy, but 26--33~per cent if the source has a
100-\micron\/ flux density $>$ 1000~Jy.  There is quite a large degree of
overlap between the $Log_{10}(S_{60}/S_{12}) \geq
1.30$~\&~$Log_{10}(S_{25}/S_{12}) \geq 0.57$ sample and the sample of
sources with a 100-\micron\/ flux density $>$ 1000~Jy.  Taking the union of
the two samples yields 17 {\em IRAS} sources within 1~arcmin, and 21
within 2~arcmin of a 6.7-GHz \methanol\/ maser, from a sample 127 {\em
IRAS} sources.  Thus there is a 13--17~per cent probability that an
{\em IRAS} source, selected using these criteria, is associated with a
6.7-GHz \methanol\/ maser.

\begin{table}
  \centering
  \caption{Distribution of 100-\micron\/ flux density for {\em IRAS} sources
	   in the region $l$ = 325\degr -- 335\degr\/, $b$ = -0\fdg53
	   -- 0\fdg53, with a moderate or higher flux quality.}
  \label{tab:100micronhist}
  \begin{tabular}{rrr}
  {\bf Log$_{10}$(S100$_{\micron}$)} & {\bf Number of } & 
    {\bf Number of} \\
				     & {\bf {\em IRAS}}       & 
    {\bf 6.7-GHz \methanol} \\ 
				     & {\bf sources}    &
    {\bf masers} \\ [2mm]
  \hline\hline
  1.0--1.5   &     0    &    0 \\
  1.5--2.0   &    11    &    0 \\
  2.0--2.5   &    69    &    1 \\
  2.5--3.0   &    52    &    2 \\
  3.0--3.5   &    22    &    4 \\
  3.5--4.0   &    12    &    5 \\
  4.0--4.5   &     7    &    2 \\
  4.5--5.0   &     1    &    0 \\
  \end{tabular}
\end{table}

\subsection{Efficiency of {\em IRAS}--based searches}

Some of the searches towards sources selected from the {\em IRAS}
Point-Source Catalog \shortcite{IRAS} detected 6.7-GHz \methanol\/
masers associated with a large fraction of those sources.  On the
other hand, this does not necessarily imply that they detected a large
fraction of the maser sources within a given region.  This survey
represents the first opportunity to examine the overall efficiency of
the various {\em IRAS} search techniques, as we are able to determine
the fraction of 6.7-GHz \methanol\/ maser sources which would be
detected by any particular search method.  Of the 41 6.7-GHz
\methanol\/ masers in the region we have surveyed, 9 have no {\em
IRAS} counterpart within 2~arcmin of the {\em IRAS} position.  We have
indicated above that we expect 3 chance positional associations
between {\em IRAS} sources and our sample of 6.7-GHz \methanol\/
masers.  In general these chance associations cannot be detected by
any criteria which exclude most of the {\em IRAS} sources.  This
implies that any {\em IRAS}-based selection criteria will not detect
more than 71~per cent of the \methanol\/ masers in any given region.
Three of the {\em IRAS} sources (15566-5304, 16159-5012 \& 16172-5028)
are within 2~arcmin of two 6.7-GHz \methanol\/ masers.  This means
that the number of 6.7-GHz masers detected by any given {\em IRAS}
selection criterion which includes any of these sources, will be
slightly greater than the number of {\em IRAS} sources with an
associated 6.7-GHz \methanol\/ maser.

The WC- criteria detect 16--21 (39--51~per cent) of the 41 6.7-GHz
\methanol\/ masers in the ``survey region'', whereas the WC criteria
detect only 14--19 (34--46~per cent).  The SWGM criteria yield the
largest fraction of {\em IRAS} sources with an associated 6.7-GHz
\methanol\/ maser, but still detect only 10--15 (24--37~per cent) of
the maser sources within the region.  Interestingly the HM criteria
yielded results slightly worse than the WC criteria, detecting less of
the 6.7-GHz \methanol\/ masers 11--17 (27--42~per cent) and selecting
a lower percentage of {\em IRAS} sources with an associated maser.
The selection criterion that the 100-\micron\/ flux density of the
{\em IRAS} source should exceed 1000 Jy also detected 11--17
(27--42~per cent) of the 6.7-GHz \methanol\/ masers in the region.
The final selection criteria we evaluated were the union of the
100-\micron\/ and WC- criteria.  A search for 6.7-GHz \methanol\/
masers towards this sample of {\em IRAS} sources would detect 17--24
(41--59~per cent) of the masers in the region.

The efficiency of the various {\em IRAS}-based searching methods, both
in terms of the fraction of {\em IRAS} sources with an associated
6.7-GHz \methanol\/ maser and the fraction of masers detected is
summarized in Table~\ref{tab:efficiency}.  It is clear that for the
various {\em IRAS} selection criteria there is a compromise between
the fraction of {\em IRAS} sources with associated 6.7-GHz \methanol\/
masers and the fraction of the total masers which are detected.  Our
analysis shows that searching toward {\em IRAS} sources with a
100-\micron\/ flux density greater than 1000~Jy is more efficient than
the WC- criteria in terms of the fraction of {\em IRAS} sources
associated with 6.7-GHz \methanol\/ masers, but detects a smaller
fraction of 6.7-GHz \methanol\/ masers.  A possible explanation for
this is that because ultra-compact \ionhy\/ regions are some of the
brightest objects in the {\em IRAS} Point-Source Catalog and their
flux density peaks near 100~\micron, many {\em IRAS} sources with a
large 100-\micron\/ flux density are likely to be UC\ionhy\/ regions.
The WC- criteria are likely to select a larger fraction of all
UC\ionhy\/ regions in the Galaxy, as they are based upon spectral
properties which do not change with distance.  This means that the WC-
criteria are likely to detect a larger fraction of the 6.7-GHz
\methanol\/ masers.  

\begin{table*}
  \begin{centering} 
  \caption{A summary of the efficiency of several {\em IRAS}-based
	   selection criteria.  Note : the lower range is for 6.7-GHz
	   \methanol\/ masers within 1~arcmin of an {\em IRAS} source,
	   and the upper for masers less than 2~arcmin from an {\em
	   IRAS} source.}
  \label{tab:efficiency}
  \end{centering}
  \begin{tabular}{lcccc}
    {\bf {\em IRAS} selection} & {\bf No. of {\em IRAS}} & {\bf No. of}       &
      {\bf Fraction of {\em IRAS}} & {\bf Fraction of the 41} \\
    {\bf method}               & {\bf candidates}        & {\bf methanol}     &
      {\bf sources yielding}       & {\bf known masers}       \\
                               &                         & {\bf sources found}&
      {\bf a detection}            & {\bf detected}           \\
  \hline\hline
  WC-                                     & 120 & 16--21 & 13--16\% & 39--51\% \\
  WC                                      &  74 & 14--19 & 19--23\% & 34--46\% \\
  SWGM                                    &  31 & 10--15 & 32--42\% & 24--37\% \\
  HM                                      &  69 & 11--17 & 16--22\% & 27--42\% \\
  $S_{100} \geq $ 1000~Jy                 &  42 & 11--17 & 26--33\% & 27--42\% \\
  Union of WC- \& $S_{100} \geq $ 1000~Jy & 127 & 17--24 & 13--17\% & 41--59\% \\
  \end{tabular}
\end{table*}

Although the {\em IRAS}-based selection criteria have been used to
detect ultra-compact \ionhy\/ regions and \methanol\/ masers, the
large beamwidth of {\em IRAS}, and the high degree of confusion in
these fields, means that an {\em IRAS} beam will typically contain
several stars in various stages of formation.  Therefore we cannot
rule out the possibility that these criteria simply select active star
formation regions containing UC\ionhy\/ regions.  In this case, the
{\em IRAS} position quoted may be severely confused, and not a good
guide to the positions of the UC\ionhy\/ regions.

\subsection{Implications of the number of UC\ionhy\/ regions in the Galaxy}

This survey has shown that a significant fraction of 6.7-GHz
\methanol\/ masers cannot be detected using any {\em IRAS}-based
selection criterion.  Whether these masers are associated with
UC\ionhy\/ regions has yet to be determined.  If they are, they may
provide a method of estimating the total number of UC\ionhy\/ regions
in the Galaxy.  Wood \& Churchwell \shortcite{Wo1989} found 1717 {\em
IRAS} sources which satisfied the criteria $Log_{10}(S_{60}/S_{12})
\geq 1.30$~\&~$Log_{10}(S_{25}/S_{12}) \geq 0.57$ and had moderate or
better quality flux density measurements at 25 and 60~\micron\/.  They
argue that the majority of these sources are UC\ionhy\/ regions.  Of
the 41 6.7-GHz \methanol\/ masers in the region we surveyed
completely, 19 at most are associated with {\em IRAS} sources which
meet these criteria.  If we assume that all 6.7-GHz \methanol\/ masers
are associated with UC\ionhy\/ regions, and that the fraction of
UC\ionhy\/ regions which have associated 6.7-GHz \methanol\/ maser
emission is the same for those which satisfy the WC- criteria as those
which do not, then this implies that the number of UC\ionhy\/ regions
in the Galaxy is more than a factor of 2 greater than that estimated
by Wood \& Churchwell.  High-resolution continuum observations have
failed to detect radio continuum emission associated with several
6.7-GHz \methanol\/ masers \cite{El1995b} which may indicate that
some masers are not associated with OB star formation (and hence
UC\ionhy\/ regions).  Alternatively, these masers may be associated
with UC\ionhy\/ regions which were too weak to detect, either because
they are very young or they are associated with a star an early B-type
star.  Our observations are consistent with an estimate of more than
3000 UC\ionhy\/ regions in the Galaxy, although the true number may be
less, depending on the validity of our assumptions.

\section{Conclusion}

The results of this present work, while not being sufficiently large
to determine parameters pertaining to the entire Galaxy, highlight a
number of interesting phenomena.  They show that many 6.7-GHz
\methanol\/ masers are not associated with sources in the {\em IRAS}
Point-Source Catalog, and some may be associated with sources that
have colours atypical of UC\ionhy\/ regions.  This may be because of
the large number of sources near the Galactic Plane, which can confuse
the {\em IRAS} flux density measurements and cause some sources to be
excluded.  We will continue to investigate these sources to determine
whether they belong to a separate population.  If these masers are
associated with UC\ionhy\/ regions, this will allow us to improve on
previous estimates of the number of such objects in the Galaxy
\cite{Wo1989}.  We have also shown that most of the {\em IRAS}-based
searches will detect less than 50~per cent of the 6.7-GHz
\methanol\/ maser sources.

\section*{Acknowledgments}

We are grateful to Phil Button and Tino Delbourgo for their work on the 
digital autocorrelation spectrometer used for these observations, to Gordon
Gowland and Phil Jenkins for assisting with some of the observing,
and to Jim Caswell for useful discussions in planning the survey.
This research has made use of the Simbad database, operated at CDS, 
Strasbourg, France.

\label{lastpage}

\end{document}